\documentclass[12pt]{article}

\begin{document}

\title{Deformation Quantization and Quaternions}
\author{Tadafumi Ohsaku}

\date{\today}

\maketitle



The concept of deformation quantization gives us a large amount of important subjects
of mathematics and theoretical/mathematical physics~[1-8].
Deformation procedures will take Poisson manifolds as their starting points of quantization~[1,2],
and the deformations are introduced perturbatively in the Poisson manifolds~[3,4,5].
It was shown that, the Kontsevich's deformation quantization formula~[3] will be converted into a 
perturbative expansion of a path integration~[5].
The conditions of Poisson manifolds $M_{p}$ 
are given by the following relations of functions $f$, $g$ and $h$ on $M_{p}$: 
(i) The commutativity, $f\cdot g=g\cdot f$, 
(ii) the associativity, $(f\cdot g)\cdot h=f\cdot (g\cdot h)$,
(iii) the realizations of the Lie-algebra relations of the Poisson brackets.
The Bayen-Flato-Fr{\o}nsdal-Lichnerowicz-Sternheimer's work gives the following definition
for the deformation quantization~[2]: 
(I) Associativity in an appropriately determined $\star$-product, namely 
$f\star(g\star h)=(f\star g)\star h$, 
(II) when a $\star$-product is expanded by a deformation parameter
$\nu$ as $f\star g=\sum_{s>0}\nu^{s}C_{s}[f,g]$, 
its coefficients of the zeroth-order $C_{0}[f,g]$ in $\nu$ becomes the usual point product $f\cdot g$,
while the first order in $\nu$ becomes
$C_{1}[f,g] = \frac{1}{2}\{f,g\}$ ( $\{\quad,\quad\}$; a Poisson bracket ).

The noncommutative field theory sometimes assume the noncommutativity of coordinate system
by $[x_{i},x_{j}]= i\theta_{ij}\ne 0$~[6,7,8], 
and these relations explicitly break associativity of algebra of the theory.
We can interpret that, 
the noncommutative field theory gives us extensions/generalizations of concepts on numbers
in the framework of quantum field theories.
Very recently, I have examined the basis of complex analysis 
under the noncommutativity $[z,\bar{z}]\ne 0$
( it is a modification of the field ${\bf C}$ ),
with the method of deformation quantization~[8].
It is a famous fact that, quaternions makes a noncommutative field of numbers.
Hence it is interesting for us to investigate a deformation quantization of functions
of quaternions.
The extension of number field might give us a new perspective on quantum theory,
similar to the cases of symmetry considerations in various areas of physics.
In this short note, we will obtain the definition of the $\star$-product of functions
of quaternions for the deformation quantization, 
and examine several algebraic properties,
especially the Poisson and the $\star$-product algebra of them.

\vspace{10mm}

First, we give the definition of quaternions.
$q$ denotes a quaternion throughout this paper:
\begin{eqnarray}
q &=& a + ib + jc + kd,  \nonumber  \\
{\bf H} &=& \{ a +ib + jc + kd | a,b,c,d\in{\bf R}\}. 
\end{eqnarray}
The quaternionic group ${\cal Q}$ is constructed by the following elements, 
given by the definitions and relations:
\begin{eqnarray}
& & {\cal Q}:\{1,-1,i,-i,j,-j,k,-k\},  \nonumber \\
-1 &=& i^{2} = j^{2} = k^{2}, \nonumber \\
(-1)^{2} &=& 1, \quad -i = (-1)i, \quad -j = (-1)j, \quad -k = (-1)k, \nonumber \\
ij &=& k, \quad ji = -k, \quad jk = i, \quad kj = -i, \quad ki = j, \quad ik = -j. 
\end{eqnarray}
Here, $1$ is the unit of the group.
The conjugate of $q$ is defined as $\bar{q} \equiv a - ib - jc - kd$,
and then one finds that the conjugation has the following relations:
\begin{eqnarray} 
& & \overline{q_{1}+q_{2}} = \bar{q_{1}}+\bar{q_{2}}, \quad \overline{q_{1}q_{2}} = \bar{q_{1}}\bar{q_{2}}, \quad 
\bar{\bar{q}} = q, \quad q = \bar{q} \; {\rm when} \; q\in{\bf R}. 
\end{eqnarray}
The norm $\|q\|$ of $q$ will be given by
\begin{eqnarray}
& & \|q\| \equiv |q| = \sqrt{a^{2}+b^{2}+c^{2}+d^{2}}, \nonumber \\
& & |q|^{2} = q\bar{q} = \bar{q}q, \quad |q| = \sqrt{q\bar{q}} = \sqrt{\bar{q}q} = |\bar{q}|, \nonumber \\
& & |q_{1}+q_{2}| \le |q_{1}| + |q_{2}|, \quad |q_{1}q_{2}| = |q_{1}| |q_{2}|.
\end{eqnarray}
Therefore, the inverse of $q$ is determined by 
\begin{eqnarray}
\frac{q\bar{q}}{|q|^{2}} &=& 1, \quad q^{-1} = \frac{\bar{q}}{|q|^{2}}. 
\end{eqnarray}
The quaternionic numbers obey the following linear algebra, the noncommutativity,
and the associativity:
\begin{eqnarray}
& & q_{1}+q_{2} = q_{2}+q_{1}, \quad
q_{1}+(q_{2}+q_{3}) = (q_{1}+q_{2})+q_{3},   \nonumber \\
& & q_{1}q_{2} \ne q_{2}q_{1}, \quad q_{1}(q_{2}q_{3}) = (q_{1}q_{2})q_{3},  \quad 
q_{1}(q_{2}+q_{3}) = q_{1}q_{2} + q_{1}q_{3}.
\end{eqnarray}
By the results of these definitions,
products of arbitrary functions $f(q),g(q),h(q);q\in{\bf H}$ have the following
noncommutativity and asscociativity:  
\begin{eqnarray}
f(q)g(q) &\ne& g(q)f(q), \quad \Bigl(f(q)g(q)\Bigr)h(q) = f(q)\Bigl(g(q)h(q)\Bigr).
\end{eqnarray}

\vspace{10mm}

Beside the noncommutativity coming from the nature of quaternions, 
it is possible for us to introduce other noncommutative relations 
( similar to the case of quantum group ),
defined between the components of $q$:
\begin{eqnarray}
&& [a,a] = [b,b] = [c,c] = [d,d] = 0, \nonumber \\
&& [a,b] = C_{ab}, \quad [a,c] = C_{ac}, \quad [a,d] = C_{ad},  \nonumber \\
&& [b,c] = C_{bc}, \quad [b,d] = C_{bd}, \quad [c,d] = C_{cd}.
\end{eqnarray}
In principle, we can take $C_{mn}$ ( $m,n$ denote the components $a,b,c,d$ of $q$ ) as quaternionic numbers
as $C_{mn} = C^{(a)}_{mn}+iC^{(b)}_{mn}+jC^{(c)}_{mn}+kC^{(d)}_{mn}$ ( $C^{(l)}_{mn}\in{\bf R}$ ).
In that case, one finds
\begin{eqnarray}
& & [q,q] = [\bar{q},\bar{q}] = 0, \nonumber \\
& & [q,\bar{q}] = -[\bar{q},q] = -2i(C_{ab}+C_{cd}) -2j(C_{ac}-C_{bd}) -2k(C_{ad}+C_{bc}).
\end{eqnarray}
If we consider these relations in our theory, this noncommutativity also have roles 
in the inequality $f(q)g(q)\ne g(q)f(q)$.
We do not consider the noncommutativity seriously in this paper,
because algebra of our theory will become very complicated and lengthy by the results of it,
and it is quite tedious for us to handle the formulae derived under the noncommutativity.

\vspace{10mm}

The Moyal-Weyl $\star$-product of two functions $f(q)$ and $g(q)$ will be introduced as follows:
\begin{eqnarray}
f(q)\star g(q) &\equiv& f(q)\exp\Bigg[\frac{\nu}{2} \Bigl(  
\overleftarrow{\partial_{a}}\Theta_{ab}\overrightarrow{\partial_{b}}
+\overleftarrow{\partial_{b}}\Theta_{ba}\overrightarrow{\partial_{a}}
+\overleftarrow{\partial_{a}}\Theta_{ac}\overrightarrow{\partial_{c}}
+\overleftarrow{\partial_{c}}\Theta_{ca}\overrightarrow{\partial_{a}}  \nonumber \\
& & \qquad
+\overleftarrow{\partial_{a}}\Theta_{ad}\overrightarrow{\partial_{d}}
+\overleftarrow{\partial_{d}}\Theta_{da}\overrightarrow{\partial_{a}}
+\overleftarrow{\partial_{b}}\Theta_{bc}\overrightarrow{\partial_{c}}
+\overleftarrow{\partial_{c}}\Theta_{cb}\overrightarrow{\partial_{b}}  \nonumber \\
& & \qquad
+\overleftarrow{\partial_{b}}\Theta_{bd}\overrightarrow{\partial_{d}}
+\overleftarrow{\partial_{d}}\Theta_{db}\overrightarrow{\partial_{b}}
+\overleftarrow{\partial_{c}}\Theta_{cd}\overrightarrow{\partial_{d}}
+\overleftarrow{\partial_{d}}\Theta_{dc}\overrightarrow{\partial_{c}} \Bigr) \Bigg]g(q). \nonumber \\
& &  
\end{eqnarray}
The deformation parameter $\nu$ and $\Theta_{mn}$ ( $m,n=a,b,c,d$ )
can be taken as quaternionic numbers, 
though the algebra of $\star$-product becomes very complicated.
We consider the case $\nu, \Theta_{mn}\in{\bf C}$.
The quantization by the deformation has been introduced into the directions 
of the Poisson brackets given by the differentiations with respect 
to real numbers $a$, $b$, $c$ and $d$, namely the components of $q$. 
For example, even if functions $f(q)$, $g(q)$ and $f(q)\cdot g(q)$ are Lorentz-invariant 
( a Lorentz transformation keeps $0=a^{2}+b^{2}+c^{2}+d^{2}=|q|^{2}$ ), 
the symmetry explicitly be broken in the $\star$-product $f(q)\star g(q)$.
We can use the operator
$\partial\equiv \partial_{a}+i\sigma_{1}\partial_{b}+i\sigma_{2}\partial_{c}+i\sigma_{3}\partial_{d}$ 
( $\sigma_{n}$ ( $n=1,2,3$ ); the Pauli matrices )
to keep the Lorentz symmetry, and can give a Moyal-Weyl-type $\star$-product
by the combinations $\partial\partial^{\dagger}$ and $\partial^{\dagger}\partial$,
though the first order in $\nu$ of the expansion of the $\star$-product will vanish identically,
because 
$\partial\partial^{\dagger}=\partial^{\dagger}\partial
=\partial^{2}_{a}+\partial^{2}_{b}+\partial^{2}_{c}+\partial^{2}_{d}$.
Under the following conditions for the tensors $\Theta_{mn}$ associated with the Poisson brackets,
\begin{eqnarray}
& & \Theta_{ab} = -\Theta_{ba}, \quad
\Theta_{ac} = -\Theta_{ca}, \quad
\Theta_{ad} = -\Theta_{da}, \nonumber \\
& & \Theta_{bc} = -\Theta_{cb}, \quad
\Theta_{bd} = -\Theta_{db}, \quad
\Theta_{cd} = -\Theta_{dc}, 
\end{eqnarray}
one finds the Poisson brackets of the $\star$-product in the canonical form:
\begin{eqnarray}
f(q)\star g(q) &=& f(q)g(q) \nonumber \\
& & + \frac{\nu}{2}\Bigg( 
\Theta_{ab}\{f,g\}^{P.B.}_{ab}+\Theta_{ac}\{f,g\}^{P.B.}_{ac}+\Theta_{ad}\{f,g\}^{P.B.}_{ad} \nonumber \\
& & \quad 
+ \Theta_{bc}\{f,g\}^{P.B.}_{bc}+\Theta_{bd}\{f,g\}^{P.B.}_{bd}+\Theta_{cd}\{f,g\}^{P.B.}_{cd} \Bigg) \nonumber \\
& & + {\cal O}(\nu^{2}).
\end{eqnarray}
Here, the Poisson brackets have been defined as follows:
\begin{eqnarray}
\{f,g\}^{P.B.}_{ab} &\equiv& 
\frac{\partial f}{\partial a}\frac{\partial g}{\partial b}
-\frac{\partial f}{\partial b}\frac{\partial g}{\partial a},  \quad
\{f,g\}^{P.B.}_{ac} \equiv 
\frac{\partial f}{\partial a}\frac{\partial g}{\partial c}
-\frac{\partial f}{\partial c}\frac{\partial g}{\partial a},  \nonumber \\
\{f,g\}^{P.B.}_{ad} &\equiv& 
\frac{\partial f}{\partial a}\frac{\partial g}{\partial d}
-\frac{\partial f}{\partial d}\frac{\partial g}{\partial a},  \quad
\{f,g\}^{P.B.}_{bc} \equiv 
\frac{\partial f}{\partial b}\frac{\partial g}{\partial c}
-\frac{\partial f}{\partial c}\frac{\partial g}{\partial b},  \nonumber \\
\{f,g\}^{P.B.}_{bd} &\equiv& 
\frac{\partial f}{\partial b}\frac{\partial g}{\partial d}
-\frac{\partial f}{\partial d}\frac{\partial g}{\partial b},  \quad
\{f,g\}^{P.B.}_{cd} \equiv 
\frac{\partial f}{\partial c}\frac{\partial g}{\partial d}
-\frac{\partial f}{\partial d}\frac{\partial g}{\partial c}.
\end{eqnarray}
If we use $q$ which has the noncommutative relations in (8),
we have to employ a kind of normal ordering of products of $a$, $b$, $c$ and $d$
similar to the situations of Grassmann numbers, as discussed in Ref.~[8]. 

\vspace{10mm}

Now, we examine the Poisson algebra and several Lie-algebraic relations of our theory.
Hereafter, we do not examine the effects of noncommutativity in components of $q$ given in (8),
and concentrate on the effect of noncommutativity coming from the quaternionic group ${\cal Q}$. 
For example, we find the following Poisson brackets by the definitions given above:
\begin{eqnarray}
0 &=& \{q,q\}^{P.B.}_{ab} = \{q,q\}^{P.B.}_{ac} = \{q,q\}^{P.B.}_{ad} \nonumber \\
&=& \{\bar{q},\bar{q}\}^{P.B.}_{ab} = \{\bar{q},\bar{q}\}^{P.B.}_{ac} = \{\bar{q},\bar{q}\}^{P.B.}_{ad} \nonumber \\
&=& \{q,\bar{q}\}^{P.B.}_{bc} = \{q,\bar{q}\}^{P.B.}_{bd} = \{q,\bar{q}\}^{P.B.}_{cd} \nonumber \\
&=& \{\bar{q},q\}^{P.B.}_{bc} = \{\bar{q},q\}^{P.B.}_{bd} = \{\bar{q},q\}^{P.B.}_{cd}, \nonumber \\
2k &=& \{q,q\}^{P.B.}_{bc} = \{\bar{q},\bar{q}\}^{P.B.}_{bc} 
= -\{q,\bar{q}\}^{P.B.}_{ad} = \{\bar{q},q\}^{P.B.}_{ad},  \nonumber \\
-2j &=& \{q,q\}^{P.B.}_{bd} = \{\bar{q},\bar{q}\}^{P.B.}_{bd} 
= \{q,\bar{q}\}^{P.B.}_{ac} = -\{\bar{q},q\}^{P.B.}_{ac},  \nonumber \\
2i &=& \{q,q\}^{P.B.}_{cd} = \{\bar{q},\bar{q}\}^{P.B.}_{cd} 
= -\{q,\bar{q}\}^{P.B.}_{ab} = \{\bar{q},q\}^{P.B.}_{ab}.
\end{eqnarray}
We have found that, the pair of $q$ and $\bar{q}$ in the Poisson brackets shows 
the symplectic structure, $\{q,\bar{q}\}^{P.B.}_{mn}=-\{\bar{q},q\}^{P.B.}_{mn}$.
We should notice that, our Poisson brackets are given in terms of the derivatives
with respect to the componets of $q$, and thus $q$ and $\bar{q}$ could not be regarded
as a pair of canonical coordinates even though the realization/appearance of the symplectic structure.
We examine other examples:
\begin{eqnarray}
\{q,q^{2}\}^{P.B.}_{ab} 
&=& \{q^{2},q\}^{P.B.}_{ab}   \nonumber \\ 
&=& -\{q,q^{2}\}^{P.B.}_{cd} +4ia
= \{q^{2},q\}^{P.B.}_{cd} -4ia  \nonumber \\
&=& -\{q,\bar{q}^{2}\}^{P.B.}_{ab} -4ia -4b
= -\{\bar{q}^{2},q\}^{P.B.}_{ab} +4ia +4b   \nonumber \\
&=& -\{q,\bar{q}^{2}\}^{P.B.}_{cd} -4ia  
= \{\bar{q}^{2},q\}^{P.B.}_{cd} +4ia \nonumber \\
&=& -\{\bar{q},q^{2}\}^{P.B.}_{ab} + 4ia -4b 
= -\{q^{2},\bar{q}\}^{P.B.}_{ab} -4ia +4b   \nonumber \\
&=& \{\bar{q},q^{2}\}^{P.B.}_{cd} + 4ia 
= -\{q^{2},\bar{q}\}^{P.B.}_{cd} -4ia   \nonumber \\
&=& \{\bar{q},\bar{q}^{2}\}^{P.B.}_{ab} 
= \{\bar{q}^{2},\bar{q}\}^{P.B.}_{ab}   \nonumber \\
&=& \{\bar{q},\bar{q}^{2}\}^{P.B.}_{cd} -4ia
= -\{\bar{q}^{2},\bar{q}\}^{P.B.}_{cd} +4ia   \nonumber \\
&=& -2b + 2ia -2iq = -2kc + 2jd, \nonumber \\
\{q,q^{2}\}^{P.B.}_{ac} 
&=& -\{q^{2},q\}^{P.B.}_{ac}   \nonumber \\ 
&=& -\{q,q^{2}\}^{P.B.}_{bd} -4ja 
= \{q^{2},q\}^{P.B.}_{bd} +4ja    \nonumber \\
&=& \{q,\bar{q}^{2}\}^{P.B.}_{ac} +4ja + 4c 
= -\{\bar{q}^{2},q\}^{P.B.}_{ac} +4ja +4c   \nonumber \\
&=& -\{q,\bar{q}^{2}\}^{P.B.}_{bd} + 4ja 
= \{\bar{q}^{2},q\}^{P.B.}_{bd} -4ja   \nonumber \\
&=& \{\bar{q},q^{2}\}^{P.B.}_{ac} -4ja +4c 
= \{q^{2},\bar{q}\}^{P.B.}_{ac} +4ja -4c   \nonumber \\
&=& \{\bar{q},q^{2}\}^{P.B.}_{bd} -4ja 
= -\{q^{2},\bar{q}\}^{P.B.}_{bd} +4ja   \nonumber \\
&=& -\{\bar{q},\bar{q}^{2}\}^{P.B.}_{ac}  
= -\{\bar{q}^{2},\bar{q}\}^{P.B.}_{ac}    \nonumber \\
&=& \{\bar{q},\bar{q}^{2}\}^{P.B.}_{bd} +4ja 
= -\{\bar{q}^{2},\bar{q}\}^{P.B.}_{bd} -4ja   \nonumber \\
&=& -2c + 2ja -2jq = -2kb + 2id, \nonumber \\
\{q,q^{2}\}^{P.B.}_{ad} 
&=& -\{q^{2},q\}^{P.B.}_{ad}   \nonumber \\
&=& \{q,q^{2}\}^{P.B.}_{bc} -4ka 
= -\{q^{2},q\}^{P.B.}_{bc} +4ka \nonumber \\
&=& \{q,\bar{q}^{2}\}^{P.B.}_{ad} + 4ka + 4d
= \{\bar{q}^{2},q\}^{P.B.}_{ad} -4ka -4d  \nonumber \\
&=& \{q,\bar{q}^{2}\}^{P.B.}_{bc} + 4ka  
= -\{\bar{q}^{2},q\}^{P.B.}_{bc} -4ka  \nonumber \\
&=& \{\bar{q},q^{2}\}^{P.B.}_{ad} -4ka +4d   
= \{q^{2},\bar{q}\}^{P.B.}_{ad} +4ka -4d   \nonumber \\
&=& -\{\bar{q},q^{2}\}^{P.B.}_{bc} -4ka  
= \{q^{2},\bar{q}\}^{P.B.}_{bc} +4ka   \nonumber \\
&=& -\{\bar{q},\bar{q}^{2}\}^{P.B.}_{ad}  
= -\{\bar{q}^{2},\bar{q}\}^{P.B.}_{ad}    \nonumber \\
&=& -\{\bar{q},\bar{q}^{2}\}^{P.B.}_{bc} +4ka  
= \{\bar{q}^{2},\bar{q}\}^{P.B.}_{bc} -4ka    \nonumber \\
&=& -2d + 2ka -2kq = 2jb - 2ic.
\end{eqnarray}
Due to the nature of quaternion, 
the symplectic structure of the Poisson manifold disappear or becomes unclear;
sometimes symplectic while other cases not. 
It has been found by the results of the Poisson algebra given above, 
the criteria of Poisson manifolds are explicitly broken in our case.
In several cases, the Jacobi identities are not satisfied because of
$\{f(q),g(q)\}^{P.B.}_{mn}\ne -\{g(q),f(q)\}^{P.B.}_{mn}$.
We obtain the following point- and $\star$- products:
\begin{eqnarray}
q\star q &=& q^{2} + \nu\Bigl( k\Theta_{bc}-j\Theta_{bd}+i\Theta_{cd} \Bigr), \nonumber \\
q\star\bar{q} &=& q\bar{q} - \nu\Bigl( i\Theta_{ab}+j\Theta_{ac}+k\Theta_{ad} \Bigr), \nonumber \\
\bar{q}\star q &=& \bar{q}q + \nu\Bigl( i\Theta_{ab}+j\Theta_{ac}+k\Theta_{ad} \Bigr), \nonumber \\
\bar{q}\star\bar{q} &=& \bar{q}^{2} + \nu\Bigl( k\Theta_{bc}-j\Theta_{bd}+i\Theta_{cd} \Bigr).
\end{eqnarray}
Hence we find that, when $\nu, \Theta_{mn}\in{\bf R}$,
$\overline{q\star q}\ne\bar{q}\star\bar{q}$ while $\overline{q\star\bar{q}}=\bar{q}\star q$.
We conclude that
\begin{eqnarray}
f(q)\star g(q) \ne g(q)\star f(q), \quad 
\overline{f(q)\star g(q)} \ne \overline{f(q)}\star\overline{g(q)}
\end{eqnarray}
in general.
Moreover,
\begin{eqnarray}
q\star q^{2} 
&=& q\star \Bigg[ q\star q - \nu\Bigl( k\Theta_{bc}-j\Theta_{bd}+i\Theta_{cd} \Bigr) \Bigg] \nonumber \\
&=& q^{3} + \nu \Bigl[ \Theta_{ab}(-kc+jd)+\Theta_{ac}(-kb+id)+\Theta_{ad}(jb-ic) \nonumber \\
& & \, +\Theta_{bc}(2ka+jb-ic)+\Theta_{bd}(-2ja+kb-id)+\Theta_{cd}(2ia+kc-jd)  \Bigr], \nonumber \\
q^{2}\star q 
&=& \Bigg[ q\star q - \nu\Bigl( k\Theta_{bc}-j\Theta_{bd}+i\Theta_{cd} \Bigr) \Bigg]\star q \nonumber \\
&=& q^{3} + \nu \Bigl[ \Theta_{ab}(-kc+jd)+\Theta_{ac}(kb-id)+\Theta_{ad}(-jb+ic) \nonumber \\
& & \, +\Theta_{bc}(2ka-jb+ic)+\Theta_{bd}(-2ja-kb+id)+\Theta_{cd}(2ia-kc+jd)  \Bigr], \nonumber \\
q\star\bar{q}^{2} 
&=& q\bar{q}^{2} + \nu \Bigl[ \Theta_{ab}(-2ia-2b+kc-jd) \nonumber \\
& & \, +\Theta_{ac}(-2ja-kb-2c+id)+\Theta_{ad}(-2ka+jb-ic-2d)   \nonumber \\
& & \, +\Theta_{bc}(-2ka+jb-ic)+\Theta_{bd}(2ja+kb-id)+\Theta_{cd}(-2ia+kc-jd)  \Bigr], \nonumber \\
\bar{q}^{2}\star q 
&=& \bar{q}^{2}q + \nu \Bigl[ \Theta_{ab}(2ia+2b+kc-jd)  \nonumber \\
& & \, +\Theta_{ac}(2ja+kb+2c-id)+\Theta_{ad}(2ka+jb-ic+2d) \nonumber \\
& & \, +\Theta_{bc}(-2ka-jb+ic)+\Theta_{bd}(2ja-kb+id)+\Theta_{cd}(-2ia-kc+jd)  \Bigr], \nonumber \\
\bar{q}\star q^{2} 
&=& \bar{q}q^{2} + \nu \Bigl[ 
\Theta_{ab}(2ia-2b+kc-jd)  \nonumber \\
& & \, +\Theta_{ac}(2ja-kb-2c+id)+\Theta_{ad}(2ka+jb-ic-2d) \nonumber \\
& & \, +\Theta_{bc}(-2ka-jb+ic)+\Theta_{bd}(2ja-kb+id)+\Theta_{cd}(-2ia-kc+jd)  \Bigr], \nonumber \\
q^{2}\star\bar{q} 
&=& q^{2}\bar{q} + \nu \Bigl[ \Theta_{ab}(-2ia+2b+kc-jd)  \nonumber \\
& & \, +\Theta_{ac}(-2ja-kb+2c+id)+\Theta_{ad}(-2ka+jd-ic+2d) \nonumber \\
& & \, +\Theta_{bc}(-2ka+jb-ic)+\Theta_{bd}(2ja+kb-id)+\Theta_{cd}(-2ia+kc-jd)  \Bigr], \nonumber \\
\bar{q}\star\bar{q}^{2} 
&=& \bar{q}^{3} + \nu \Bigl[ \Theta_{ab}(-kc+jd)+\Theta_{ac}(kb-id)+\Theta_{ad}(-jb+ic) \nonumber \\
& & \, +\Theta_{bc}(2ka-jb+ic)+\Theta_{bd}(-2ja-kb+id)+\Theta_{cd}(2ia-kc+jd)  \Bigr], \nonumber \\
\bar{q}^{2}\star\bar{q} 
&=& \bar{q}^{3} + \nu \Bigl[ \Theta_{ab}(-kc+jd) +\Theta_{ac}(kb-id)+\Theta_{ad}(-jb+ic) \nonumber \\
& & \, +\Theta_{bc}(2ka+jb-ic)+\Theta_{bd}(-2ja+kb-id)+\Theta_{cd}(2ia+kc-jd)  \Bigr]. \nonumber \\
\end{eqnarray}
Clearly, the structures of these $\star$-products reflect the quaternionic group ${\cal Q}$.
Here, we observe $q^{2}\star q\ne q\star q^{2}$, etc.
From these results, one finds
$(q\star q)\star q - q\star(q\star q) \ne 0$, 
$(q\star q)\star \bar{q} - q\star(q\star \bar{q}) \ne 0$,
$(q\star \bar{q})\star q - q\star(\bar{q}\star q) \ne 0$, etc.
The associativity is broken, and we conclude that,
\begin{eqnarray}
\Bigl[ f(q)\star g(q) \Bigr] \star h(q) \ne f(q)\star \Bigl[ g(q)\star h(q) \Bigr].
\end{eqnarray}

In summary, we have shown various conditions and criterions 
for the Poisson manifold and the deformation quantization are broken 
by the nature of quaternion.

\vspace{10mm}

Now, final comments are in order. 
It is well-known fact that, by using appropriate definitions of matrix notations for $q$, 
we can regard ${\bf H}={\bf R}^{4}$ or ${\bf H}={\bf C}^{2}$.
It is interesting for us to investigate the relations between our theory and
noncommutative field theories of ${\bf R}^{4}$ or ${\bf C}^{2}$.
It is also an important problem for us to investigate the path integral expansion/interpretation
for the $\star$-product of our theory.
It is now in progress to investigate the quaternionic $\star$-product by the methods
of topological field/string ( the Poisson $\sigma$-model ) theories.

\end{document}